\documentclass[aps,preprint]{revtex4}
\usepackage{amssymb}
\usepackage{amsmath}

\input{tcilatex}

\begin{document}

\title{One-loop Massive Scattering Amplitudes and Ward Identities in String
Theory}
\author{Chuan-Tsung Chan}
\email{ctchan@phys.cts.nthu.edu.tw}
\affiliation{Physics Division, National Center for Theoretical Sciences, Hsinchu, Taiwan,
R.O.C.}
\author{Jen-Chi Lee}
\email{jcclee@cc.nctu.edu.tw}
\affiliation{Department of Electrophysics, National Chiao-Tung University, Hsinchu,
Taiwan, R.O.C.}
\date{\today }

\begin{abstract}
We calculate bosonic open string one-loop massive scattering amplitudes for
some low-lying string states. By using the periodicity relations of Jacobi
theta functions, we explicitly prove an infinite number of one-loop type I
stringy Ward identities derived from type I zero-norm states in the old
covariant first quantized (OCFQ) spectrum of open bosonic string. The
subtlety in the proofs of one-loop type II stringy Ward identities is
discussed by comparing them with those of string-tree cases. High-energy
limit of these stringy Ward identities can be used to fix the
proportionality constants between one-loop massive high-energy scattering
amplitudes of different string states with the same momenta. These
proportionality constants \textit{can not} be calculated directly from
sample calculation as we did previously in the cases of string-tree
scattering amplitudes.
\end{abstract}

\maketitle

\section{Introduction}

Recently it was discovered that \cite{1,2} the high-energy limit $\alpha
^{\prime }\rightarrow \infty $ of stringy Ward identities, or massive gauge
invariances, derived from the decoupling of two types of zero-norm states
imply an infinite number of linear relations \cite{3} among high energy
scattering amplitudes of different string states with the same momenta. The
calculation was first done for mass levels $m^{2}=4,6$ and was soon
generalized to arbitrary mass levels \cite{4,5}. These linear relations can
be used to fix the proportionality constants between high energy scattering
amplitudes of different string states algebraically at each fixed mass
level. These proportionality constants were found to be independent of the
scattering angle $\phi _{CM}$ and the loop order $\chi $ of string
perturbation theory. Thus there is only one independent component of
high-energy string scattering amplitudes for each fixed mass level. For the
case of string-tree amplitudes, a general formula can even be given to
express all high-energy stringy scattering amplitudes at arbitrary mass
levels in terms of those of tachyons \cite{1,6}. Other approaches of stringy
symmetries can be found in \cite{7,8}.

The importance of zero-norm states and their implication on stringy
symmetries were first pointed out in the context of massive $\sigma $-model
approach of string theory \cite{9}. On the other hand, zero-norm states were
also shown \cite{10} to carry the spacetime $\omega _{\infty }$ symmetry
charges of 2D string theory. Some implications of stringy Ward identities on
the scattering amplitudes were also discussed in \cite{11}. All the above
zero-norm state calculations are independent of the high-energy saddle-point
calculations of Gross and Mende \cite{12}, Gross \cite{3} and Gross and
Manes \cite{13}. In fact, the results of saddle-point calculations by those
authors were found \cite{1,2,6} to be inconsistent with stringy Ward
identities, which are valid to all energy, and thus could threat the
validity of unitarity of string perturbation theory. A corrected
saddle-point calculation was given in Ref [6] where the missing terms of the
calculation in Ref [3,12,13] were identified to recover the stringy Ward
identities.

In this paper, we shall first calculate bosonic open string one-loop massive
scattering amplitudes for some low-lying string states which were not
calculated in the literature. Bosonic open string tree massive scattering
amplitudes were calculated in \cite{11}, and the tree-level massive gauge
invariances were explicitly justified for the first few mass levels. General
tree-level gauge invariances were proved by "the canceled propagator
argument" in the operator approach in \cite{14}. On the other hand, bosonic
closed string one-loop massless scattering amplitudes were calculated in 
\cite{15}, and the one-loop modular invariance was justified there. Here we
are aiming to explicitly show the one-loop massive gauge invariances or Ward
identities. High-energy limit of these stringy Ward identities can be used
to fix the proportionality constants between one-loop massive high-energy
scattering amplitudes of different string states with the same momenta.

Unlike the string-tree scattering amplitudes, which can be exactly
integrated to calculate their high-energy limit, one-loop scattering
amplitudes are not exactly integrable and their high-energy limit are
difficult to calculate. Thus the determination of the proportionality
constants between high-energy one-loop scattering amplitudes relies solely
on the algebraic high-energy stringy Ward identities, and can not be
calculated directly from sample calculation as we did previously \cite{1,2}
in the cases of string-tree scattering amplitudes. This is one of the main
motivation to explicitly prove one-loop stringy Ward identities in this
paper. In section II of this paper, we first give a new proof of tree-level
stringy Ward identities, which will be useful for the proof of one-loop Ward
identities in section III. In section III, by using the periodicity
relations of Jacobi theta functions, we will show an infinite number of type
I one-loop stringy Ward identities derived from type I zero-norm states in
the old covariant first quantized (OCFQ) spectrum of open bosonic string.
The subtlety in the proofs of one-loop type II stringy Ward identities will
be discussed by comparing them with those of string-tree cases. The explicit
proof of type II one-loop Ward identities seem to be much more involved and
are, presumably, related to more advanced identities of Jacobi theta
functions. In section IV, high-energy limit of stringy Ward identities will
be used to fix the proportionality constants between one-loop massive
high-energy scattering amplitudes of different string states. These
proportionality constants are otherwise difficult to calculate directly from
sample calculation as in the cases of string-tree scattering amplitudes. We
thus have explicitly justify Gross's conjecture \cite{3}, for the first
time, that the proportionality constants between high energy scattering
amplitudes are independent of the scattering angle $\phi _{CM}$\ and the
loop order $\chi $\ of string perturbation theory, at least for $\chi =1,0$.
A brief conclusion is given in section V.

\section{ A new proof of tree-level stringy Ward identities}

For illustration and setting up the notations, let's begin with simple
examples of string tree-level massive scattering amplitudes of the first
massive level. For the string-tree level $\chi =1$, with one tensor $v_{2}$
and three tachyons $v_{1,3,4}$, all scattering amplitudes of mass level $%
m^{2}=2$ were calculated in \cite{11}. These are

\begin{eqnarray}
\mathcal{T}^{\mu \nu } &=&\int \prod_{i=1}^{4}dx_{i}<e^{ik_{1}X}\partial
X^{\mu }\partial X^{\nu }e^{ik_{2}X}e^{ik_{3}X}e^{ik_{4}X}>  \TCItag{1} \\
&=&\frac{\Gamma (-\frac{s}{2}-1)\Gamma (-\frac{t}{2}-1)}{\Gamma (\frac{u}{2}%
+2)}[t/2(t/2+1)k_{1}^{\mu }k_{1}^{\nu }  \notag \\
&&-2(s/2+1)(t/2+1)k_{1}^{(\mu }k_{3}^{\nu )}+s/2(s/2+1)k_{3}^{\mu
}k_{3}^{\nu }],  \TCItag{2}
\end{eqnarray}%
\begin{eqnarray}
\mathcal{T}^{\mu } &=&\int \prod_{i=1}^{4}dx_{i}<e^{ik_{1}X}\partial
^{2}X^{\mu }e^{ik_{2}X}e^{ik_{3}X}e^{ik_{4}X}>  \TCItag{3} \\
&=&\frac{\Gamma (-\frac{s}{2}-1)\Gamma (-\frac{t}{2}-1)}{\Gamma (\frac{u}{2}%
+2)}[-t/2({t}/2+1)k_{1}^{\mu }-s/2({s}/2+1)k_{3}^{\mu }],  \TCItag{4}
\end{eqnarray}%
where $s=-(k_{1}+k_{2})^{2},t=-(k_{2}+k_{3})^{2},$ and $u=-(k_{1}+k_{3})^{2}$
are the Mandelstam variables. In deriving Eqs.(2) and (4), we have made the $%
SL(2,R)$ gauge fixing and restricted to the $s-t$ channel of the amplitudes
by choosing $x_{1}=0,0\leqq x_{2}\leqq 1,x_{3}=1,x_{4}=\infty .$

In the OCFQ spectrum of open bosonic string theory, the solutions of
physical states conditions include positive-norm propagating states and two
types of zero-norm states. The latter are \cite{14} 
\begin{eqnarray}
\text{Type I}:L_{-1}\left\vert x\right\rangle , &&\text{ where }%
L_{1}\left\vert x\right\rangle =L_{2}\left\vert x\right\rangle =0,\text{ }%
L_{0}\left\vert x\right\rangle =0;  \TCItag{5} \\
\text{Type II}:(L_{-2}+\frac{3}{2}L_{-1}^{2})\left\vert \widetilde{x}%
\right\rangle , &&\text{ where }L_{1}\left\vert \widetilde{x}\right\rangle
=L_{2}\left\vert \widetilde{x}\right\rangle =0,\text{ }(L_{0}+1)\left\vert 
\widetilde{x}\right\rangle =0.  \TCItag{6}
\end{eqnarray}%
Eqs.(5) and (6) can be derived from Kac determinant in conformal field
theory. While type I states have zero-norm at any spacetime dimension, type
II states have zero-norm \textit{only} at D=26. In the first quantized
approach of string theory, the stringy \textit{on-shell} Ward identities are
proposed to be \cite{11} (for simplicity we choose four-point amplitudes in
this paper) 
\begin{equation}
\mathcal{T}_{\chi }(k_{i})=g_{c}^{2-\chi }\int \frac{Dg_{\alpha \beta }}{%
\mathcal{N}}DX^{\mu }\exp (-\frac{\alpha ^{\prime }}{2\pi }\int d^{2}\xi 
\sqrt{g}g^{\alpha \beta }\partial _{\alpha }X^{\mu }\partial _{\beta }X_{\mu
})\overset{4}{\underset{i=1}{\Pi }}v_{i}(k_{i})=0,  \tag{7}
\end{equation}%
where at least one of the 4 vertex operators corresponds to the zero-norm
state solution of Eqs.(5) or (6). In Eq.(7) $g_{c}$ is the closedstring
coupling constant, $\mathcal{N}$ is the volume of the group of
diffeomorphisms and Weyl rescalings of the worldsheet metric, and $%
v_{i}(k_{i})$ are the on-shell vertex operators with momenta $k_{i}$. The
integral is over orientable open surfaces of Euler number $\chi $
parametrized by moduli $\overrightarrow{m}$ with punctures at $\xi _{i}$.
For the first massive level, $m^{2}=2$, there are two zero-norm states, and
the corresponding string tree-level $\chi =1$ Ward identities were
explicitly calculated to be \cite{11} 
\begin{eqnarray}
k_{\mu }\theta _{\nu }\mathcal{T}^{\mu \nu }+\theta _{\mu }\mathcal{T}^{\mu
} &=&0,  \TCItag{8} \\
(\frac{3}{2}k_{\mu }k_{\nu }+\frac{1}{2}\eta _{\mu \nu })\mathcal{T}^{\mu
\nu }+\frac{5}{2}k_{\mu }\mathcal{T}^{\mu } &=&0,  \TCItag{9}
\end{eqnarray}%
where $\theta _{\nu }$ is a transverse vector. In Eqs.(8) and (9), we have
chosen, say, $v_{2}(k_{2})$ to be the vertex operators constructed from
zero-norm states and $k_{\mu }\equiv k_{2\mu }$. Note that Eq.(8) is the
type I Ward identity while Eq.(9) is the type II Ward identity which is
valid only at $D=26$.

The proof of the decoupling theorem at string-tree amplitudes for general
mass levels, without explicit calculations of massive scattering amplitudes,
has been demonstrated in \cite{14}, where cyclic symmetry is used to show
that both types of zero-norm states decouple from the on-shell correlation
functions. Unfortunately, both approaches in \cite{7} and \cite{10} only
work for string tree amplitudes, and one can not extend similar arguments to
stringy amplitudes at loop levels. For this reason, it is instructive to
give a new proof of the decoupling of zero-norm states for the string-tree
amplitudes. As we will see soon that the essential features of this new
proof will be maintained in our proof for the decoupling theorem at one-loop
level. Also this new proof illustrates some subtle features assoicated with
the proof of the decoupling theorem for type II zero-norm states.

Our stragedy for proving the decopling theorem is to rewrite the stringy
amplitudes as an integral of worldsheet total derivatives, and the boundary
terms vanish due to the special properties of the string propagator. Taking
the massless state in open bosonic string theory as an example 
\begin{equation}
\mathcal{T}^{\mu }\equiv \int \prod_{i=1}^{4}dx_{i}<e^{ik_{1}X}\partial
X^{\mu }e^{ik_{2}X}e^{ik_{3}X}e^{ik_{4}X}>,  \tag{10}
\end{equation}%
\begin{eqnarray}
k_{2\mu }\mathcal{T}^{\mu } &=&\int_{0}^{1}dx\mbox{ }x^{(1,2)}(1-x)^{(2,3)}[%
\frac{(1,2)}{x}-\frac{(2,3)}{1-x}]  \notag \\
&=&\int_{0}^{1}dx\mbox{ }\frac{\partial }{\partial x}[x^{(1,2)}(1-x)^{(2,3)}]
\notag \\
&=&[x^{(1,2)}(1-x)^{(2,3)}]|_{0}^{1}=0.  \TCItag{11}
\end{eqnarray}%
Notice that we have introduced a convention for inner products among
external momenta, e.g., $(1,2)\equiv k_{1}\cdot k_{2}$. This derivation has
employed the $SL(2,R)$ gauge fixing to reduce the four point integrations
into a single integral, and we list only the $s-t$ channel of the scattering
amplitude. However, it should be easy to generalize our derivation to the
case without doing this $SL(2,R)$ fixing. In particular, for the one-loop
open string amplitudes the residual gauge symmetry is a $U(1)$ symmetry,
thus we can not fix the positions of vertex operators (three out of four) as
in the case of tree amplitudes. Still, as we shall show later, the total
derivative argument can be applied at the one-loop level, at least in the
case of type I zero norm state.

The fact that scatterig amplitudes containing a vertex operator of the
massless zero-norm state can be expressed as an integral of total derivative
should come as no suprise, since the vertex operator for $m^{2}=0$ zero-norm
state can be written as a worldsheet total derivative, 
\begin{equation}
v(k,\zeta =k)=k\cdot \partial X\exp ^{ikX}=-i\partial (\exp ^{ikX}), 
\tag{12}
\end{equation}%
where the partial derivative means derivative with respect to the worldsheet
time variable. Indeed, according to eq.(5), all type I zero-norm states are
generated by the $L_{-1}$ Virasoro generator, which is a partial derivative
on the holomorphoic coordinate of string worldsheet.

To illustrate this point further for the cases of massive scattering
amplitudes, let us work out the case for type I singlet zero-norm state at $%
m^{2}=4$. The stringy Ward identity associated with this state is, 
\begin{equation}
(\frac{17}{4}k_{\mu }k_{\nu }k_{\lambda }+\frac{9}{2}k_{\mu }\eta _{\nu
\lambda })\mathcal{T}^{(\mu \nu \lambda )}+(21k_{\mu }k_{\nu }+9\eta _{\mu
\nu })\mathcal{T}^{(\mu \nu )}+25k_{\mu }\mathcal{T}{\ }^{\mu }=0,  \tag{13}
\end{equation}%
where we have defined the $m^{2}=4$ scattering amplitudes as 
\begin{eqnarray}
\mathcal{T}^{\mu \nu \lambda } &\equiv &\int
\prod_{i=1}^{4}dx_{i}<e^{ik_{1}X}\partial X^{\mu }\partial X^{\nu }\partial
X^{\lambda }e^{ik_{2}X}e^{ik_{3}X}e^{ik_{4}X}>,  \TCItag{14} \\
\mathcal{T}^{(\mu \nu )} &\equiv &\int
\prod_{i=1}^{4}dx_{i}<e^{ik_{1}X}\partial X^{(\mu }\partial ^{2}X^{\nu
)}e^{ik_{2}X}e^{ik_{3}X}e^{ik_{4}X}>,  \TCItag{15} \\
\mathcal{T}^{\mu } &\equiv &\int \prod_{i=1}^{4}dx_{i}<e^{ik_{1}X}\partial
^{3}X^{\mu }e^{ik_{2}X}e^{ik_{3}X}e^{ik_{4}X}>.  \TCItag{16}
\end{eqnarray}%
After some algebraic manipulations, we can rewrite the sum of the left hand
side of Eq.(13) as an integral 
\begin{eqnarray}
\int_{0}^{1}dx &&[\frac{17}{4}\frac{\partial ^{3}}{\partial x^{3}}M_{0,0}-%
\frac{33}{4}\frac{\partial ^{2}}{\partial x^{2}}M_{1,0}+\frac{33}{4}\frac{%
\partial ^{2}}{\partial x^{2}}M_{0,1}+\frac{3}{4}((1,2)+(2,3))\frac{\partial 
}{\partial x}M_{1,1}  \notag \\
&+&\frac{3}{4}\frac{\partial }{\partial x}M_{2,0}-9\frac{\partial }{\partial
x}M_{1,1}+\frac{3}{4}\frac{\partial }{\partial x}M_{0,2}]=0.  \TCItag{17}
\end{eqnarray}%
Here we have introduced a notation for the factor in the stringy amplitudes 
\begin{equation}
M_{p,q}\equiv x^{(1,2)-p}(1-x)^{(2,3)-q}.  \tag{18}
\end{equation}%
Based on these observations and the fact that all type I zero-norm states
can be written as a worldsheet total derivative, one can easily extend our
proof to all type I zero-norm states, and conclude that all stringy
scattering amplitudes for a type I zero-norm state and any other physical
states can be written as integrals of worldsheet total derivatives and thus
vanish due to the boundary conditions.

Having shown that one can use the total derivative argument to prove
decoupling theorem for type I zero-norm states, one natural question is that
whether we can use the same trick to prove the decoupling theorem for type
II zero-norm states. As we have emphasized before \cite{1,2}, the decoupling
of type II zero-norm states is of crucial importance for demonstrating
linear relations between stringy scattering amplitudes at high energies.
Nevertheless, given the definition of type II zero-norm states, Eq.(6), it
can be shown that, in general, type II zero-norm states can not be written
as worldsheet total derivatives, thus our proof for type I zero-norm states
seems need to be modified. Fortunately, a detailed investigation shows that,
at least in the case of tree amplitudes, one can still express the stringy
amplitudes associated with type II zero-norm states as integrals of total
derivatives, and the boundary terms vanish as before. For instance, at first
massive level, $-k^{2}=m^{2}=2$, we have one singlet type II zero-norm
state, 
\begin{equation}
\lbrack (\frac{3}{2}k_{\mu }k_{\nu }+\frac{1}{2}\eta _{\mu \nu })\alpha
_{-1}^{\mu }\alpha _{-1}^{\nu }+\frac{5}{2}k_{\mu }\alpha _{-2}^{\mu }]|0,k>.
\tag{19}
\end{equation}%
The amplitude of this state with three tachyons can be written as 
\begin{eqnarray}
&&\int_{0}^{1}dx\{\frac{3}{2}\frac{\partial ^{2}}{\partial x^{2}}M_{0,0}+%
\frac{\partial }{\partial x}[M_{0,1}-M_{1,0}]\}  \notag \\
&=&\frac{3}{2}\frac{\partial }{\partial x}%
M_{0,0}|_{0}^{1}+[M_{0,1}-M_{1,0}]|_{0}^{1}=0.  \TCItag{20}
\end{eqnarray}%
At the second massive level, $m^{2}=4$, we have two vector zero-norm states, 
$D_{1}$ and $D_{2}$, which we chose to be linear combinations of the
original type I and type II vector zero-norm states, and the stringy Ward
identities associated with them are \cite{11} 
\begin{eqnarray}
&(\frac{5}{2}k_{\mu }k_{\nu }\theta _{\lambda }+\eta _{\mu \nu }\theta
_{\lambda })\mathcal{T}^{(\mu \nu \lambda )}+9k_{\mu }\theta _{\nu }\mathcal{%
T}^{(\mu \nu )}+6\theta _{\mu }\mathcal{T}^{\mu }=&0,  \TCItag{21} \\
&(\frac{1}{2}k_{\mu }k_{\nu }\theta _{\lambda }^{\prime }+2\eta _{\mu \nu
}\theta _{\lambda }^{\prime })\mathcal{T}^{(\mu \nu \lambda )}+9k_{\mu
}\theta _{\nu }^{\prime }\mathcal{T}^{[\mu \nu ]}-6\theta _{\mu }^{\prime }%
\mathcal{T}^{\mu }=&0.  \TCItag{22}
\end{eqnarray}%
The amplitudes of these states with three tachyons can be written as 
\begin{eqnarray}
&&\int_{0}^{1}dx\left\{ (\theta \cdot k_{1})[\frac{5}{2}\frac{\partial ^{2}}{%
\partial x^{2}}M_{1,0}-\frac{3}{2}\frac{\partial }{\partial x}M_{2,0}+2\frac{%
\partial }{\partial x}M_{1,1}]\right.   \notag \\
&+&(\theta \cdot k_{3})[-\frac{5}{2}\frac{\partial ^{2}}{\partial x^{2}}%
M_{0,1}-\frac{3}{2}\frac{\partial }{\partial x}M_{0,2}+2\frac{\partial }{%
\partial x}M_{1,1}]\left. {}\right\} =0,  \TCItag{23} \\
&&\int_{0}^{1}dx\left\{ (\theta ^{\prime }\cdot k_{1})[\frac{1}{2}\frac{%
\partial ^{2}}{\partial x^{2}}M_{1,0}+\frac{3}{2}\frac{\partial }{\partial x}%
M_{2,0}+4\frac{\partial }{\partial x}M_{1,1}]\right.   \notag \\
&+&(\theta ^{\prime }\cdot k_{3})[-\frac{1}{2}\frac{\partial ^{2}}{\partial
x^{2}}M_{0,1}+\frac{3}{2}\frac{\partial }{\partial x}M_{0,2}+4\frac{\partial 
}{\partial x}M_{1,1}]\left. {}\right\} =0.  \TCItag{24}
\end{eqnarray}%
Notice that in these derivations, one needs to use momentum conservation and
on-shell conditions for vertex operators. Instead of scattering with three
tachyons, the derivation here can be generalized to arbitrary three string
states.

In the next section, we shall extend the total derivative argument of our
proof for the string-tree amplitudes in this section to the string one-loop
amplitudes, where subtlety arises for the calculation of type II zero-norm
Ward identities.

\section{One-loop stringy Ward identities}

It is believed that the decoupling of zero-norm states in string theory, or
the stringy Ward identities, should hold true for all loop orders in string
perturbation theory. Nevertheless, a mathematical proof of this assertion is
non-existent and, as we will see soon, our investigation shows some subtlies
associated with the proof of the decopling theorem for type II zero-norm
states. To begin with, we first discuss the decoupling theorem of type I
zero-norm state at one-loop level, the stringy amplitude for one massless
zero-norm state scatters with three tachyons is calculated to be 
\begin{eqnarray}
\mathcal{T}_{\mu } &=&g^{4}\int \prod_{i=1}^{4}dx_{i}<\partial X_{\mu
}e^{ik_{1}X}e^{ik_{2}X}e^{ik_{3}X}e^{ik_{4}X}>,  \TCItag{25} \\
k_{1}^{\mu }\cdot \mathcal{T}_{\mu } &=&-g^{4}\int_{0}^{1}\frac{d\omega }{%
\omega ^{2}}\int_{\rho _{2}}^{1}\frac{d\rho _{1}}{\rho _{1}}\int_{\rho
_{3}}^{1}\frac{d\rho _{2}}{\rho _{2}}\int_{\omega }^{1}\frac{d\rho _{3}}{%
\rho _{3}}{f(\omega )}^{-24}(\frac{-2\pi }{\ln \omega })^{13}  \notag \\
&&\psi _{12}^{(1,2)}\psi _{13}^{(1,3)}\psi _{14}^{(1,4)}\psi
_{23}^{(2,3)}\psi _{24}^{(2,4)}\psi _{34}^{(3,4)}[(1,2)\eta _{12}+(1,3)\eta
_{13}+(1,4)\eta _{14}],  \TCItag{26}
\end{eqnarray}%
where $g$ is the open string coupling constant. Here we follow the notations
of Green, Schwarz and Witten \cite{10}, and the one-loop open string
propagator is given by%
\begin{equation}
\ln \psi _{rs}\equiv \ln \psi (c_{sr},\omega )=\langle X^{\mu }(\rho
_{r})X^{\nu }(\rho _{s})\rangle ,  \tag{27}
\end{equation}%
where%
\begin{equation}
\ln \psi (c,\omega )=-\frac{1}{2}\ln c+\frac{\ln ^{2}c}{2\ln \omega }%
-\sum_{m=1}^{\infty }\frac{c^{m}+(\omega /c)^{m}-2\omega ^{m}}{m(1-\omega
^{m})}.  \tag{28}
\end{equation}%
Here $\rho _{r},\omega $, and $c_{sr}$ are related to the worldsheet time
coordinates $\tau _{1},\tau _{2}$, etc, as follows 
\begin{equation}
\rho _{r}\equiv e^{-(\tau _{1}+\tau _{2}+....+\tau _{r})},r=1,2,3,4;\hspace{%
0.5cm}\omega \equiv \rho _{4};\hspace{0.5cm}c_{sr}\equiv \frac{\rho _{s}}{%
\rho _{r}}.  \tag{29}
\end{equation}%
In Eq.(28), the $\psi $ function can be recasted in terms of the Jacobi $%
\theta $ function 
\begin{eqnarray}
\psi (\rho ,\omega ) &=&\frac{1-\rho }{\sqrt{\rho }}\exp (\frac{\ln ^{2}\rho 
}{2\ln \omega })\prod_{n=1}^{\infty }\frac{(1-\omega ^{n}\rho )(1-{\omega
^{n}}/\rho )}{(1-\omega ^{n})^{2}}  \TCItag{30} \\
&=&-2\pi ie^{-i\pi {\nu ^{2}}/\tau }\frac{\theta _{1}(-\nu /\tau )|-1/\tau )%
}{\theta _{1}^{\prime }(0|-1/\tau )},  \TCItag{31}
\end{eqnarray}%
where the Jacobi $\theta _{1}$ function satisfies the important periodicity
relations 
\begin{eqnarray}
\theta _{1}(\nu +1|\tau ) &=&-\theta _{1}(\nu |\tau ),  \TCItag{32} \\
\theta _{1}(\nu +\tau |\tau ) &=&-e^{-i\pi \tau -2i\pi \nu }\theta _{1}(\nu
|\tau ),  \TCItag{33}
\end{eqnarray}%
and $\nu $ and $\tau $ are defined to be 
\begin{equation}
\nu \equiv \frac{\ln \rho }{\ln \omega },\text{ \ }\tau \equiv -\frac{2\pi i%
}{\ln \omega }.  \tag{34}
\end{equation}%
For the calculations of massive scattering amplitudes, we also need the
following expressions which can be obtained by taking higher derivatives of
one-loop string propagator $\ln \psi $ in Eq.(28) 
\begin{eqnarray}
\eta (c_{rs},\omega ) &=&\langle \frac{\partial }{\partial \tau _{r}}X^{\mu
}(\rho _{r})X^{\mu }(\rho _{s})\rangle   \TCItag{35} \\
&=&c_{sr}\frac{\partial }{\partial c_{sr}}\ln \psi (c_{sr},\omega ), 
\TCItag{36}
\end{eqnarray}%
where%
\begin{equation}
\eta (c,\omega )=-\frac{1}{2}+(\frac{\ln c}{\ln \omega })-\frac{c}{1-c}%
+\sum_{n=1}^{\infty }\left( \frac{{\omega ^{n}}/c}{1-{\omega ^{n}}/c}-\frac{c%
{\omega ^{n}}}{1-c{\omega ^{n}}}\right) ;  \tag{37}
\end{equation}%
and%
\begin{eqnarray}
\Omega (c_{rs},\omega ) &=&\langle -\frac{\partial ^{2}}{\partial \tau
_{r}^{2}}X^{\mu }(\rho _{r})X^{\nu }(\rho _{s})\rangle   \TCItag{38} \\
&=&-c_{sr}\frac{\partial }{\partial c_{sr}}\eta (c_{sr},\omega ), 
\TCItag{39}
\end{eqnarray}%
where%
\begin{equation}
\Omega (c,\omega )=-(\frac{1}{\ln \omega })+\frac{c}{(1-c)^{2}}%
+\sum_{n=1}^{\infty }\left( \frac{{\omega ^{n}}/c}{(1-{\omega ^{n}}/c)^{2}}+%
\frac{c{\omega ^{n}}}{(1-c{\omega ^{n}})^{2}}\right) .  \tag{40}
\end{equation}%
\qquad 

Due to the residual conformal symmetry $U(1)$ of the one-loop open string
worldsheet, in addition to the moduli parameter $\omega $, there are three
points of vertex operators we need to integrate over for a four-point
scattering amplitude, and we have chosen the first vertex operator
corresponding to the zero-norm state. To calculate the one-loop Ward
identity for the massless zero-norm state, we make the following
observation. Taking the first term in the square bracket of Eq.(26) as an
example, we can rewrite 
\begin{eqnarray}
(1,2)\psi _{12}^{(1,2)}\eta _{12} &=&(1,2)\psi _{12}^{(1,2)-1}\frac{\partial 
}{\partial \ln c_{21}}\psi _{12}  \notag \\
&=&\frac{\partial }{\partial \ln c_{21}}\psi _{12}^{(1,2)}  \notag \\
&=&-\frac{\partial }{\partial \ln \rho _{1}}\psi _{12}^{(1,2)}=-\rho _{1}%
\frac{\partial }{\partial \rho _{1}}\psi _{12}^{(1,2)}.  \TCItag{41}
\end{eqnarray}%
Similarly, the next two terms in the square bracket of Eq.(26) can be
rewritten as a partial derivative acting on $\psi _{13}^{(1,3)}$ and $\psi
_{14}^{(1,4)}$ with respect to the variable $\rho _{1}$. Put them all
together, we get 
\begin{eqnarray}
k_{1}^{\mu }\cdot \mathcal{T}_{\mu } &=&g^{4}\int_{0}^{1}\frac{d\omega }{%
\omega ^{2}}\int_{\rho _{2}}^{1}\frac{d\rho _{1}}{\rho _{1}}\int_{\rho
_{3}}^{1}\frac{d\rho _{2}}{\rho _{2}}\int_{\omega }^{1}\frac{d\rho _{3}}{%
\rho _{3}}{f(\omega )}^{-24}(\frac{-2\pi }{\ln \omega })^{13}  \notag \\
&&\rho _{1}\frac{\partial }{\partial \rho _{1}}[\psi _{12}^{(1,2)}\psi
_{13}^{(1,3)}\psi _{14}^{(1,4)}\psi _{23}^{(2,3)}\psi _{24}^{(2,4)}\psi
_{34}^{(3,4)}].  \TCItag{42}
\end{eqnarray}%
Now, we can perform the integration by parts for the variable $\rho _{1}$,
and rewrite the integral as two surface terms 
\begin{eqnarray}
k_{1}^{\mu }\cdot \mathcal{T}_{\mu } &=&g^{4}\int_{0}^{1}\frac{d\omega }{%
\omega ^{2}}\int_{\rho _{3}}^{1}\frac{d\rho _{2}}{\rho _{2}}\int_{\omega
}^{1}\frac{d\rho _{3}}{\rho _{3}}{f(\omega )}^{-24}(\frac{-2\pi }{\ln \omega 
})^{13}  \notag \\
&&[\psi _{12}^{(1,2)}\psi _{13}^{(1,3)}\psi _{14}^{(1,4)}\psi
_{23}^{(2,3)}\psi _{24}^{(2,4)}\psi _{34}^{(3,4)}]|_{\rho _{2}}^{1}. 
\TCItag{43}
\end{eqnarray}%
However, both terms vanish due to the periodicity properties of the $\psi $
function, or Eqs.(32)-(33) of the Jacobi $\theta _{1}$ function. For the
upper limit $\rho _{1}=1$, we have 
\begin{equation}
\psi _{14}|_{\rho _{1}=1}=\psi (\frac{\omega }{\rho _{1}},\omega )|_{\rho
_{1}=1}=\psi (\omega ,\omega )=\psi (1,\omega )=0.  \tag{44}
\end{equation}%
On the other hand, for the lower limit $\rho _{1}=\rho _{2}$, we have 
\begin{equation}
\psi _{12}|_{\rho _{1}=\rho _{2}}=\psi (\frac{\rho _{2}}{\rho _{1}},\omega
)|_{\rho _{1}=\rho _{2}}=\psi (1,\omega )=0.  \tag{45}
\end{equation}%
Again, we have assumed that both $(1,2)\sim -\frac{s}{2}$ and $(1,4)\sim -%
\frac{t}{2}$ are positive numbers, and have extended the decoupling theorem
to physical region via analytical continuation, as we have done in the proof
for tree amplitudes.

The Ward identity, Eqs.(43)-(45), corresponding to the massless zero-norm
state serves as a typical example of one-loop decoupling theorem for type I
zero-norm state. We can follow the similar procedure and calculate the
one-loop Ward identity for $m^{2}=2$ vector zero-norm state 
\begin{equation}
(\theta _{\mu }k_{1\nu }\alpha _{-1}^{\mu }\alpha _{-1}^{\nu }+\theta _{\mu
}\alpha _{-2}^{\mu })|0,k_{1}>,\hspace{1cm}-k_{1}^{2}=m^{2}=2.  \tag{46}
\end{equation}%
First of all, we define the following one-loop amplitudes 
\begin{eqnarray}
\mathcal{T}^{\mu \nu } &\equiv &g^{4}\int \prod_{i=1}^{4}dx_{i}<\partial
X^{\mu }\partial X^{\nu }e^{ik_{1}X}e^{ik_{2}X}e^{ik_{3}X}e^{ik_{4}X}>, 
\TCItag{47} \\
\mathcal{T}^{\mu } &\equiv &g^{4}\int \prod_{i=1}^{4}dx_{i}<\partial
^{2}X^{\mu }e^{ik_{1}X}e^{ik_{2}X}e^{ik_{3}X}e^{ik_{4}X}>.  \TCItag{48}
\end{eqnarray}%
These amplitudes are calculated to be 
\begin{eqnarray}
\theta ^{\mu }k_{1}^{\nu }\mathcal{T}_{\mu \nu } &=&g^{4}\int_{0}^{1}\frac{%
d\omega }{\omega ^{2}}\int_{\rho _{2}}^{1}\frac{d\rho _{1}}{\rho _{1}}%
\int_{\rho _{3}}^{1}\frac{d\rho _{2}}{\rho _{2}}\int_{\omega }^{1}\frac{%
d\rho _{3}}{\rho _{3}}{f(\omega )}^{-24}(\frac{-2\pi }{\ln \omega })^{13} 
\notag \\
&\times &\psi _{12}^{(1,2)}\psi _{13}^{(1,3)}\psi _{14}^{(1,4)}\psi
_{23}^{(2,3)}\psi _{24}^{(2,4)}\psi _{34}^{(3,4)}  \notag \\
&\times &\left\{ (\theta \cdot k_{2})[(1,2)\eta _{12}^{2}+(1,3)\eta
_{12}\eta _{13}+(1,4)\eta _{12}\eta _{14}]\right.   \notag \\
&+&(\theta \cdot k_{3})[(1,3)\eta _{13}^{2}+(1,2)\eta _{12}\eta
_{13}+(1,4)\eta _{13}\eta _{14}]  \notag \\
&+&\left. (\theta \cdot k_{4})[(1,4)\eta _{14}^{2}+(1,2)\eta _{12}\eta
_{14}+(1,3)\eta _{13}\eta _{14}]\right\} ,  \TCItag{49} \\
&&  \notag \\
\theta ^{\mu }\cdot \mathcal{T}_{\mu } &=&-g^{4}\int_{0}^{1}\frac{d\omega }{%
\omega ^{2}}\int_{\rho _{2}}^{1}\frac{d\rho _{1}}{\rho _{1}}\int_{\rho
_{3}}^{1}\frac{d\rho _{2}}{\rho _{2}}\int_{\omega }^{1}\frac{d\rho _{3}}{%
\rho _{3}}{f(\omega )}^{-24}(\frac{-2\pi }{\ln \omega })^{13}  \notag \\
&\times &\psi _{12}^{(1,2)}\psi _{13}^{(1,3)}\psi _{14}^{(1,4)}\psi
_{23}^{(2,3)}\psi _{24}^{(2,4)}\psi _{34}^{(3,4)}  \notag \\
&\times &[(\theta \cdot k_{2})\Omega _{12}+(\theta \cdot k_{3})\Omega
_{13}+(\theta \cdot k_{4})\Omega _{14}].  \TCItag{50}
\end{eqnarray}%
Based on the same trick in Eqs.(41)-(45), we can now combine these results
to obtain 
\begin{eqnarray}
\theta ^{\mu }k_{1}^{\nu }\mathcal{T}_{\mu \nu }+\theta ^{\mu }\mathcal{T}%
_{\mu } &=&g^{4}\int_{0}^{1}\frac{d\omega }{\omega ^{2}}\int_{\rho _{2}}^{1}%
\frac{d\rho _{1}}{\rho _{1}}\int_{\rho _{3}}^{1}\frac{d\rho _{2}}{\rho _{2}}%
\int_{\omega }^{1}\frac{d\rho _{3}}{\rho _{3}}{f(\omega )}^{-24}(\frac{-2\pi 
}{\ln \omega })^{13}  \notag \\
&\times &\frac{\partial }{\partial \ln \rho _{1}}\{\psi _{12}^{(1,2)}\psi
_{13}^{(1,3)}\psi _{14}^{(1,4)}\psi _{23}^{(2,3)}\psi _{24}^{(2,4)}\psi
_{34}^{(3,4)}  \notag \\
&&\times \lbrack (\theta \cdot k_{2})\eta _{12}+(\theta \cdot k_{3})\eta
_{13}+(\theta \cdot k_{4})\eta _{14}]\}.  \TCItag{51}
\end{eqnarray}%
It can then be shown that, upon integration by parts, Eq.(51) vanishes due
to the periodicity properties Eqs.(32)-(33) of the Jacobi $\theta _{1}$
function.

To calculate the one-loop stringy Ward identity for the type II zero-norm
state at $m^{2}=2$, we first decompose the combination of stringy amplitudes
into two terms%
\begin{equation}
(\frac{3}{2}k_{\mu }k_{\nu }+\frac{1}{2}\eta _{\mu \nu })\mathcal{T}^{\mu
\nu }+\frac{5}{2}k_{\mu }\mathcal{T}^{\mu }=\frac{3}{2}[k_{\mu }k_{\nu }%
\mathcal{T}^{\mu \nu }+k_{\mu }\mathcal{T}^{\mu }]+[\frac{1}{2}\eta _{\mu
\nu }\mathcal{T}^{\mu \nu }+k_{\mu }\mathcal{T}^{\mu }].  \tag{52}
\end{equation}%
The first term in the decomposition can be expressed as an integral of a
worldsheet total derivative as following 
\begin{eqnarray}
(I) &\equiv &\frac{3}{2}[k_{\mu }k_{\nu }\mathcal{T}^{\mu \nu }+k_{\mu }%
\mathcal{T}^{\mu }]  \notag \\
&=&\frac{3}{2}g^{4}\int_{0}^{1}\frac{d\omega }{\omega ^{2}}\int_{\rho
_{2}}^{1}\frac{d\rho _{1}}{\rho _{1}}\int_{\rho _{3}}^{1}\frac{d\rho _{2}}{%
\rho _{2}}\int_{\omega }^{1}\frac{d\rho _{3}}{\rho _{3}}{f(\omega )}^{-24}(%
\frac{-2\pi }{\ln \omega })^{13}  \notag \\
&&\times \psi _{12}^{(1,2)}\psi _{13}^{(1,3)}\psi _{14}^{(1,4)}\psi
_{23}^{(2,3)}\psi _{24}^{(2,4)}\psi _{34}^{(3,4)}  \notag \\
&&\times \left\{ (1,2)^{2}\eta _{12}^{2}+(1,3)^{2}\eta
_{13}^{2}+(1,4)^{2}\eta _{14}^{2}\right.   \notag \\
&+&2(1,2)(1,3)\eta _{12}\eta _{13}+2(1,2)(1,4)\eta _{12}\eta
_{14}+2(1,3)(1,4)\eta _{13}\eta _{14}  \notag \\
&&-\left. (1,2)\Omega _{1,2}-(1,3)\Omega _{1,3}-(1,4)\Omega _{1,4}\right\}  
\notag \\
&=&\frac{3}{2}g^{4}\int_{0}^{1}\frac{d\omega }{\omega ^{2}}\int_{\rho
_{2}}^{1}\frac{d\rho _{1}}{\rho _{1}}\int_{\rho _{3}}^{1}\frac{d\rho _{2}}{%
\rho _{2}}\int_{\omega }^{1}\frac{d\rho _{3}}{\rho _{3}}{f(\omega )}^{-24}(%
\frac{-2\pi }{\ln \omega })^{13}  \notag \\
&&\times \frac{\partial ^{2}}{\partial \ln \rho _{1}^{2}}[\psi
_{12}^{(1,2)}\psi _{13}^{(1,3)}\psi _{14}^{(1,4)}\psi _{23}^{(2,3)}\psi
_{24}^{(2,4)}\psi _{34}^{(3,4)}].  \TCItag{53}
\end{eqnarray}%
The second term on the right hand side of Eq.(52) can be further decomposed
into two pieces

\begin{eqnarray}
(II) &\equiv &\frac{1}{2}\eta _{\mu \nu }\mathcal{T}^{\mu \nu }+k_{\mu }%
\mathcal{T}^{\mu }  \notag \\
&=&g^{4}\int_{0}^{1}\frac{d\omega }{\omega ^{2}}\int_{\rho _{2}}^{1}\frac{%
d\rho _{1}}{\rho _{1}}\int_{\rho _{3}}^{1}\frac{d\rho _{2}}{\rho _{2}}%
\int_{\omega }^{1}\frac{d\rho _{3}}{\rho _{3}}{f(\omega )}^{-24}(\frac{-2\pi 
}{\ln \omega })^{13}  \notag \\
&&\times \lbrack \psi _{12}^{(1,2)}\psi _{13}^{(1,3)}\psi _{14}^{(1,4)}\psi
_{23}^{(2,3)}\psi _{24}^{(2,4)}\psi _{34}^{(3,4)}]  \notag \\
&&\times \left\{ \frac{k_{2}^{2}}{2}\eta _{12}^{2}+\frac{k_{3}^{2}}{2}\eta
_{13}^{2}+\frac{k_{4}^{2}}{2}\eta _{14}^{2}+(2,3)\eta _{12}\eta
_{13}+(2,4)\eta _{12}\eta _{14}+(3,4)\eta _{13}\eta _{14}\right.   \notag \\
&&-\left. (1,2)\Omega _{1,2}-(1,3)\Omega _{1,3}-(1,4)\Omega _{1,4}\right\}  
\TCItag{54} \\
&=&g^{4}\int_{0}^{1}\frac{d\omega }{\omega ^{2}}\int_{\rho _{2}}^{1}\frac{%
d\rho _{1}}{\rho _{1}}\int_{\rho _{3}}^{1}\frac{d\rho _{2}}{\rho _{2}}%
\int_{\omega }^{1}\frac{d\rho _{3}}{\rho _{3}}{f(\omega )}^{-24}(\frac{-2\pi 
}{\ln \omega })^{13}  \notag \\
&&\times \frac{\partial }{\partial \ln \rho _{1}}\left\{ \psi
_{12}^{(1,2)}\psi _{13}^{(1,3)}\psi _{14}^{(1,4)}\psi _{23}^{(2,3)}\psi
_{24}^{(2,4)}\psi _{34}^{(3,4)}[\eta _{1,2}+\eta _{1,3}+\eta _{1,4}]\right\}
+\Delta ,  \TCItag{55}
\end{eqnarray}%
where the extra piece $\Delta $ in Eq.(55) is given by 
\begin{eqnarray}
\Delta  &\equiv &g^{4}\int_{0}^{1}\frac{d\omega }{\omega ^{2}}\int_{\rho
_{2}}^{1}\frac{d\rho _{1}}{\rho _{1}}\int_{\rho _{3}}^{1}\frac{d\rho _{2}}{%
\rho _{2}}\int_{\omega }^{1}\frac{d\rho _{3}}{\rho _{3}}{f(\omega )}^{-24}(%
\frac{-2\pi }{\ln \omega })^{13}  \notag \\
&&\times \lbrack \psi _{12}^{(1,2)}\psi _{13}^{(1,3)}\psi _{14}^{(1,4)}\psi
_{23}^{(2,3)}\psi _{24}^{(2,4)}\psi _{34}^{(3,4)}]  \notag \\
&&\times \left\{ (1+(1,2))(\eta _{12}^{2}-\Omega _{12})+(1+(1,3))(\eta
_{13}^{2}-\Omega _{13})+(1+(1,4))(\eta _{14}^{2}-\Omega _{14})\right\} . 
\TCItag{56}
\end{eqnarray}%
From the expressions above, one sees that while piece $(I)$ can be written
as an integral of a worldsheet total derivative, piece $(II)$ fails to be an
integral of a worldsheet total derivative. Consequently, it seems that we
can not apply the integration by parts to show the total amplitudes to be
zero.

One might be curious about the difference between string-tree and one-loop
calculations, and wonders why the simple total derivative argument can not
be applied at one-loop level. To see this, we replace the one-loop string
propagator $\ln \psi $ by the tree level propagator $\ln (x_{1}-x_{2})$ in
Eq.(56). The derivatives of the one-loop string propagator become $\eta =%
\frac{1}{x_{1}-x_{2}}$ and $\Omega =\frac{1}{(x_{1}-x_{2})^{2}}$
respectively. After making these replacements, the extra terms in piece $(II)
$, which are propotional to $\eta ^{2}-\Omega $, vanish identically.

Another observation is that in our proofs of type I one-loop Ward
identities, the vanishings of Eqs.(43) and (51) are valid for all values of
moduli parameter $\omega $. That means one need not do the $\omega $
integration to prove the type I Ward identities. On the contrary, it may
happen that an explicit $\omega $ integration is needed in order to prove
the type II Ward identities. If this is the case, the proof of \textit{closed%
} string type II Ward identities will be closely related to the $SL(2,Z)$
modular invariance of one-loop massive scattering amplitudes on torus.
Although the proofs of modular invariance of bosonic closed string one-loop 
\textit{massless} scattering amplitudes were given in \cite{15} as we have
mentioned in section I, the proofs for the \textit{massive} cases are still
lacking. It is thus of interest to see the relations between type II
one-loop Ward identities and the one-loop modular invariance of massive
stringy scattering amplitudes.

\section{High energy limit of stringy Ward identities}

The proof of stringy Ward identities for both types of zero-norm states at
one-loop level is not only of interest for demonstrating one-loop stringy
gauge invariances, but also important in extracting the high energy behavior
of one-loop scattering amplitudes. As we have shown in the previous works 
\cite{1,2} that, by taking high energy limit of these Ward identities, one
can obtain linear relations among scattering amplitudes of different string
states with the same momenta. In particular, we can exactly determine the
proportionality constants between high energy scattering amplitudes of
different string states algebraically without any integration.

For the cases of string-tree scattering amplitudes, which can be exactly
integrated to calculate their high-energy limits, these proportionality
constants have been explicitly calculated through a set of sample
calculations and were found to agree with the algebraic calculations based
on the stringy Ward identities. For the cases of string one-loop scattering
amplitudes, however, exact integrations for the sample calculations are not
possible and their high-energy limit are difficult to calculate. Thus the
determination of the proportionality constants between high-energy one-loop
scattering amplitudes relies solely on the algebraic high-energy stringy
Ward identities. This is one of the main motivation to explicitly prove
one-loop stringy Ward identities in this paper.

To illustrate this point in a more concrete way, we shall use the simplest
case as an example. We first discuss the string-tree case. At the first
massive level $m^{2}=2$, we have two string-tree Ward identities, Eqs.(8)
and (9), which have been proved by various methods discussed in section II.
In order to take high energy limits of these Ward identities, we need to
define the following orthonormal polarization vectors for the second string
vertex $v_{2}(k_{2})$ 
\begin{eqnarray}
e_{P} &=&\frac{1}{m_{2}}(E_{2},\mathrm{k}_{2},0)=\frac{k_{2}}{m_{2}}, 
\TCItag{57} \\
e_{L} &=&\frac{1}{m_{2}}(\mathrm{k}_{2},E_{2},0),  \TCItag{58} \\
e_{T} &=&(0,0,1)  \TCItag{59}
\end{eqnarray}%
in the CM frame contained in the plane of scattering. In the high energy
limit, we define the following projections of stringy amplitudes 
\begin{eqnarray}
\mathcal{T}^{\alpha \beta } &\equiv &e_{\mu }^{\alpha }e_{\nu }^{\beta
}\cdot \mathcal{T}^{\mu \nu },\hspace{2cm}\alpha ,\beta =P,L,T,  \TCItag{60}
\\
\mathcal{T}^{\alpha } &\equiv &e_{\mu }^{\alpha }\cdot \mathcal{T}^{\mu },%
\hspace{2.7cm}\alpha ,\beta =P,L,T,  \TCItag{61}
\end{eqnarray}%
where $\mathcal{T}^{\alpha \beta }$ and $\mathcal{T}^{\alpha }$ are defined
similarly as in Eqs.(1) and (3), except that $v_{1,3,4}$ can now be any
string vertex and we have conventionally put the zero-norm state at the
second vertex. After taking the high energy limit of the stringy Ward
identities and identifying $\mathcal{T}^{\cdot \cdot P\cdot \cdot }=\mathcal{%
T}^{\cdot \cdot L\cdot \cdot }$ in Eq.(9) \cite{6} , Eqs.(8) and (9) reduce
to 
\begin{eqnarray}
\sqrt{2}\mathcal{T}_{TP}^{3\rightarrow 1}+\mathcal{T}_{T}^{1} &=&0, 
\TCItag{62} \\
\sqrt{2}\mathcal{T}_{LL}^{4\rightarrow 2}+\mathcal{T}_{L}^{2} &=&0, 
\TCItag{63} \\
6\mathcal{T}_{LL}^{4\rightarrow 2}+\mathcal{T}_{TT}^{2}+5\sqrt{2}\mathcal{T}%
_{L}^{2} &=&0.  \TCItag{64}
\end{eqnarray}%
In the above equations, we have denoted the naive power counting for orders
in energy \cite{1,2} in the superscript of each amplitude according to the
following rules, $e_{L}\cdot k\sim E^{2},e_{T}\cdot k\sim E^{1}$. Note that
since $\mathcal{T}_{TP}^{1}$ is of subleading order in energy, in general $%
\mathcal{T}_{TP}^{1}\neq \mathcal{T}_{TL}^{1}$. A simple calculation of
Eqs.(62)-(64) shows that 
\begin{equation}
\mathcal{T}_{TP}^{1}:\mathcal{T}_{T}^{1}=1:-\sqrt{2},  \tag{65}
\end{equation}%
\begin{equation}
\mathcal{T}_{TT}^{2}:\mathcal{T}_{LL}^{2}:\mathcal{T}_{L}^{2}\text{ }=\text{
\ }4:1:-\sqrt{2}.  \tag{66}
\end{equation}%
It is interesting to see that, in addition to the leading order amplitudes
in Eq.(66), the subleading order amplitudes in Eq.(65) are also proportional
to each other. This does not seem to happen at higher mass level \cite{2}.
Since the proportionality constants in Eqs.(65)-(66) are independent of
particles chosen for vertex $v_{1,3,4}$, we will choose them, for example,
to be tachyons to do the sample calculation. The string-tree level
calculations by both energy expansion method \cite{1,2} and the saddle-point
method \cite{6} give the same results%
\begin{eqnarray}
\mathcal{T}_{T}^{1} &=&4E^{5}\sin \phi _{CM}\mathcal{T}(2)=-\sqrt{2}\mathcal{%
T}_{TP}^{1},  \TCItag{67} \\
\mathcal{T}_{TT}^{2} &=&4E^{6}\sin ^{2}\phi _{CM}\mathcal{T}(2)=4\mathcal{T}%
_{LL}^{2}=-2\sqrt{2}\mathcal{T}_{L}^{2}\text{ ,}  \TCItag{68}
\end{eqnarray}%
where $\mathcal{T}(2)=-\frac{1}{4}\sqrt{\pi }E^{-5}(\sin \frac{\phi _{CM}}{2}%
)^{-3}(\cos \frac{\phi _{CM}}{2})\exp (-\frac{s\ln s+t\ln t-(s+t)\ln (s+t)}{2%
}).$ Eqs.(67) and (68) agree with Eqs.(65) and (66) respectively as
expected. We have also checked that $\mathcal{T}_{TP}^{1}\neq \mathcal{T}%
_{TL}^{1}$.

We now discuss the string one-loop case. The calculations of Eqs.(57)-(66)
go through except that we have no sample calculations, Eqs.(67) and (68),
for the string one-loop case. This is due to the fact that exact
integrations, apart from numerical calculations, for the one-loop
amplitudes, Eqs.(49) and (50), are not possible and their high-energy limit
are difficult to calculate. Also, the string one-loop saddle-point
calculations of Refs \cite{3,12,13} are not reliable \cite{1,2,6}. For
example, the calculation of Ref \cite{13} predicts $\mathcal{T}_{LL}^{2}$ to
be of the subleading order in energy compared to $\mathcal{T}_{TT}^{2}$,
which obviously violates the high-energy stringy Ward identity Eq.(66). 
\textit{Thus, the one-loop stringy Ward identities calculations become, so
far, the only way to determine the one-loop proportionality constants in
Eqs.(65) and (66).} This is one of the main motivation to prove one-loop
stringy Ward identities in this paper as we have stressed before.

Notice that Eq.(65) is obtained from the high energy one-loop Ward identity
for type I zero-norm state, Eq.(51), which has been proved in the previous
section. The validity of Eq.(66), or Eqs.(63) and (64), however, relies on
the proof of high energy one-loop Ward identities for both Type I and Type
II zero-norm states, Eqs.(51) and (52). Unfortunately, we are not able to
explicitly prove Eq.(52) at this moment. Thus, strictly speaking, only
Eq.(65) is rigously proved at one-loop level but not Eq.(66) .\textit{\ }%
Eq.(65) is our first example to explicitly justify Gross's conjecture \cite%
{3} that the proportionality constants between high energy scattering
amplitudes are independent of the scattering angle $\phi _{CM}$\ and the
loop order $\chi $\ of string perturbation theory, at least for $\chi =1,0$.%
\textit{\ }Eq.(65) is remarkable in the sense that although both $T_{TL}^{1}$%
\ and $T_{T}^{1}$\ are not one-loop exactly calculable, they are indeed
proportional to each other and the ratio of them is determined by the
one-loop stringy Ward identity, Eq.(51). While a complete proof of one-loop
decoupling theorem for type II zero-norm states might require sophiscated
use of non-trivial identities of Jacobi theta functions, the validity of
these type II stringy Ward identities should be a resonable consequence from
stringy gauge symmetries and the unitarity of the theory. Thus, even though
one can not exactly integrate Eqs.(49) and (50), we do believe that these
one-loop scattering amplitudes are proportional to each other in the high
energy limit, and the proportionality constants can be determined exactly by
simple algebric means. This simple example of one-loop $m^{2}=2$ amplitudes
calculations serve as an illustrative example for the power of zero-norm
state approach \cite{1,2,6}, and can be generalized to higher massive levels
and higher genus amplitudes.

\section{Summary and Conclusion}

In this paper, we have studied one-loop massive scattering amplitudes and
their associated Ward identities in bosonic open string theory. A new proof
of the decoupling of two types of zero-norm states at string-tree level is
given which allows us to express the scattering amplitudes containing
zero-norm states as integrals of worldsheet total derivatives. Based on the
explicit one-loop calculations of four-point scattering amplitudes for some
low-lying massive string states, we show that the same technique for proving
string-tree level Ward identities can be generalized to the case of type I
zero-norm states. However, the one-loop Ward identities for type II
zero-norm states can not be proved in the same way. The subtlety in the
proofs of one-loop type II stringy Ward identities are discussed by
comparing them with those of string-tree cases. Finally, as an example,
high-energy limit of $m^{2}=2$ stringy Ward identities are used to fix the
proportionality constants between one-loop massive high-energy scattering
amplitudes at mass level $m^{2}=2$. It is interesting to see that, in
addition to the leading order amplitudes, the subleading order amplitudes
are also proportional to each other. This does not seem to happen at higher
mass level. These proportionality constants can not be calculated directly
from sample calculations as we did in the cases of string-tree scattering
amplitudes.

It should be clear from our study in this paper that the explicit proof of
the decoupling theorem for type II zero-norm states at one-loop level is of
crucial importance. Presumably, one needs some higher identities of Jacobi
theta functions.

\section{Acknowledgments}

This work is supported in part by the National Science Council, Taiwan,
R.O.C. We would like to thank Pei-Ming Ho for many valuable discussions. We
would also like to thank NCTS/TPE for the hospitality.

\end{document}